\documentclass[%
mathleft,%
final,%
]{an}
\usepackage{graphicx}
\usepackage{amsmath}
\usepackage{times}
\usepackage{color}
\usepackage{nicefrac}
\newcommand{\kms}{km\,s$^{-1}$}
\newcommand{\ms}{m\,s$^{-1}$}
\newcommand{\epseri}{$\epsilon$\,Eri}
\newcommand{\Bf}{$\mathrm{B\cdot \sqrt{f}}$}
\newcommand{\B}{$\langle B \rangle$}
\newcommand{\vect}[1]{\mathbf{#1}}

\begin{document}

\Pagespan{1}{}
\Yearpublication{2015}%
\Yearsubmission{2014}%
\Month{3}%
\Volume{336}%
\Issue{3}%
\DOI{10.1002/asna.201412162}%

\title{Magnetic field measurements of $\epsilon$~Eridani from Zeeman broadening\thanks{Based on data obtained with the STELLA robotic telescopes in Tenerife, an AIP facility jointly operated with IAC.}}

\author{L. T. Lehmann, A. K\"unstler, T. A. Carroll, \and K. G. Strassmeier\thanks{Corresponding author.
\email{kstrassmeier@aip.de}}}
\authorrunning{L. T. Lehmann et al.}
\institute{Leibniz-Institut f\"ur Astrophysik Potsdam (AIP), An der
Sternwarte 16, D-14482 Potsdam, Germany}

\received{2014}
\accepted{2015}
\publonline{2015}


\abstract{%
We present new magnetic field measurements of the K2 main-sequence star \epseri\ based on principal components analysis (PCA) line-profile reconstructions. The aim of this paper is to quantify the surface-averaged magnetic field and search for possible variations. A total of 338 optical echelle spectra from our robotic telescope facility STELLA with a spectral resolution of 55,000 were available for analysis. This time-series was used to search for the small line-profile variations due to a surface magnetic field with the help of a PCA. Evidence for a spatial and temporal inhomogeneous magnetic field distribution is presented. 
The mean, surface averaged, magnetic field strength was found to be $\langle B \rangle = 186 \pm 47$\,G in good agreement with previous Zeeman-broadening measurements. Clear short-term variations of the surface averaged magnetic field of up to few tens Gauss were detected together with evidence for a three-year cycle in the surface-averaged magnetic field of \epseri.
}

\maketitle


\section{Introduction}

\epseri\ is among the bright stars that are mildly magnetically active and were a planet detection had been claimed. Low-amplitude radial velocities variations of \epseri\ were first reported by Campbell et al. (\cite{camp}). \cite{Hatzes2000} had announced a planet around \epseri\ with an orbital period of 2500\,d and an eccentricity of 0.6. Benedict et al. (\cite{ben:mca}) and  Reffert \& Quirrenbach (\cite{ref:qui}) refined the orbital solution and combined the radial velocities with astrometric observations and derived a likely mass for the planet of 1.55\,M$_{\rm Jup}$. However, Zechmeister et al. (\cite{zech}) could not confirm the existence of this planet from higher-precision radial-velocity data nor did Anglada-Escud\'e \& Butler (\cite{ang:but}) succeed to coherently combine the different radial velocity data sets with the original 2500-d period.

Detailed studies of the basic parameters of \epseri\ were derived from high-resolution spectroscopy by Steenbock \& Holweger (\cite{ste:hol}), Drake \& Smith (\cite{dra:smi}), Valenti et al. (\cite{val}), R\"uedi et al. (\cite{ruedi}), among others, and from optical interferometry by Baines \& Armstrong (\cite{bai:arm}). These studies converged on the picture that \epseri\ is a single, moderately-young ($\approx$\,800\,Myr), K2 main-sequence star with an effective temperature of 5100\,K, a mass of 0.82\,M$_\odot$, a radius of 0.74\,R$_\odot$, subsolar metallicity of $-0.17$, and barely resolvable rotational line broadening ($v\sin i$\,=\,1.6\,\kms ). The stellar rotation period was first determined from the Mt.~Wilson H\&K data to 11.10\,d by Gray \& Baliunas (\cite{gra:bal}), its range further discussed in Donahue et al. (\cite{don:saa}). Time-series photometry was obtained by Frey et al. (\cite{frey}) and from the 12 spots detected with a rotation period of between $\approx$\,10\,--\,12\,d a differential rotation very much equal to the Sun's was derived. Croll et al. (\cite{cro:wal}), Fr\"ohlich (\cite{fro}), and Savanov (\cite{sav}) analyzed photometric data from the MOST satellite and found two spots or spot groups separated by 180$^\circ$ in longitude and individual spot rotation periods of 11.35\,d and 11.55\,d from which a differential surface rotation of roughly half the solar value was derived.

Gray \& Baliunas (\cite{gra:bal}) found evidence of a 5-year activity cycle from Mt.~Wilson H\&K observations. The cycle period was revised just recently by \cite{Metcalfe2013} from a 45-yr long Ca\,{\sc ii} H\&K data set based on Mt.~Wilson, CTIO, Lowell, CASLEO, CPS, and HARPS observations. They found two coexisting 2.95$\pm$0.03\,yr and 12.7$\pm$0.3\,yr periods with evidence for a Maunder minimum-like state in the late 1980's followed by the resurgence of a coherent 3-year cycle.

%
%
%
\begin{figure*}
\includegraphics[trim=40 0 10 0, angle=0,width=\textwidth,clip]{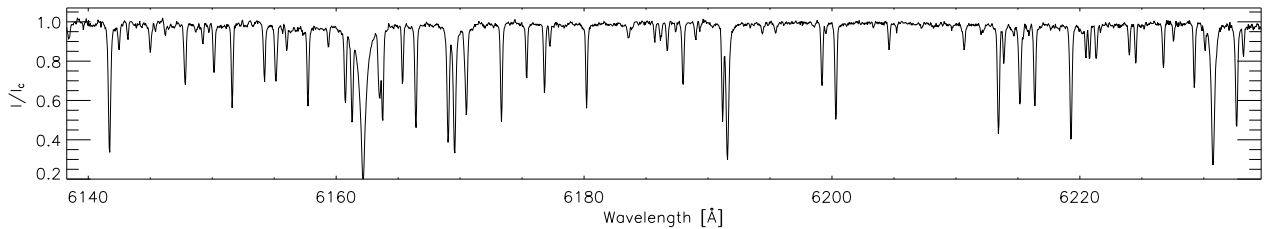}
\caption{STELLA example spectrum of \epseri . Shown is the full echelle order \#\,92 from 6140--6230\,\AA\ containing the magnetic-sensitive Fe\,{\sc i} line at 6173\,\AA . The spectral resolution is 55,000.}\label{F1}
\end{figure*}

The line broadening of \epseri\ due to a magnetic field was first announced by Marcy (\cite{marcy}) and Gray (\cite{gray84}). Marcy (\cite{marcy}) inferred a surface magnetic field between 620\,G and 2850\,G and a filling factor, $f$, between 0.89 and 0.2, respectively. These values were confined to 1\,kG and $f=0.35$ in subsequent work by Basri \& Marcy (\cite{bas:mar}) and Mary \& Basri (\cite{mar:bas}). Gray (\cite{gray84}) obtained a magnetic field strength of 1.9\,kG and $f=0.3$. Saar (\cite{saar}) improved the detections by consequently applying radiative-transfer techniques in magnetized atmospheres but still obtained rather different magnetic-field values between 0 and 3.5\,kG from different spectral lines. He, among others, pointed out the limitation due to substantial systematic effects in the data, in particular due to line blending. First infrared  magnetic-field measurements from Zeeman broadening were presented by Valenti et al. (\cite{val}) and yielded a photospheric coverage of just 8.8\,\%\ with a 1.44\,kG field. R\"uedi et al. (\cite{ruedi}) re-discussed the impact of blending and confined the spatially averaged surface field strength on \epseri\ to 165$\pm$30\,G and concluded that it is difficult (at best) to separate $f$ and $B$. Finally, Kochukhov et al. (\cite{koch}) presented modern linear and circular polarization observations with the new HARPS spectropolarimeter. The Stokes-V profiles of \epseri\ changed systematically from night to night throughout the rotation period of the star. The mean longitudinal magnetic field varied between $-5.8\pm 0.1$\,G and $+4.7\pm 0.1$\,G. Recently, \cite{Jeffers2014} could also detect variations in the large-scale magnetic field geometry. They examined the evolution of the magnetic field topology of \epseri\ using spectropolarimetric observations and Zeeman-Doppler imaging for six observational epochs spread over a period of nearly seven years. They found similar S-index variation for the last seven years as \cite{Metcalfe2013} including the hint of a three year cycle. Using Zeeman-Doppler imaging \cite{Jeffers2014} showed a highly variable magnetic field geometry. The poloidal component varies from strong dipolar to monopolar and the toroidal component varies from being non-existent to being the dominant component of the magnetic field energy.   

In this paper, we present and analyze time-series optical spectroscopy of $\epsilon$\,Eri. We employed our robotic telescope STELLA-II and its echelle spectrograph SES in Tenerife. Section~\ref{S2} presents and describes the new data, Sect.~\ref{S3} describes the Zeeman-broadening PCA technique and Sect.~\ref{S4} presents the results. Our conclusions are presented in Sect.~\ref{S5}.

\section{Observations and data reductions}\label{S2}

Echelle spectroscopy was taken with the 1.2\,m STELLA telescopes between July 2008 and October 2013. Spectra were taken with an exposure time of 300\,s and
achieved a signal-to-noise (S/N) ratio between 400:1 to 100:1 per resolution element depending on weather conditions. A total of 338 spectra were used in this analysis.
STELLA-I and STELLA-II are fully robotic telescope that make up the STELLA observatory at the Iz\~{a}na ridge on Tenerife in the Canary
islands (Strassmeier et al.~\cite{stella}, \cite{malaga}). The
fiber-fed STELLA Echelle Spectrograph (SES) is the one telescope's
only instrument. It is a white-pupil spectrograph with an R2 grating
with two off-axis collimators, a prism cross disperser and a
folded Schmidt camera with an E2V 2k$\times$2k CCD as the
detector (until mid 2012). All spectra have a fixed format on the CCD and cover the
wavelength range from 388--882\,nm with increasing inter-order gaps
near the red end starting at 734\,nm towards 882\,nm. The resolving
power is $R$=55,000 corresponding to a spectral resolution of
0.12~\AA \ at 650\,nm (3-pixel sampling). An example spectrum is
shown in Fig.~\ref{F1}. We note that the SES received a major
upgrade in summer 2012 with a new cross disperser, a new optical
refractive camera, and a new 4k$\times$4k CCD. A bit earlier, the SES fiber was moved to
the prime focus of the second STELLA telescope in 2011. Further
details of the performance of the system were reported by Weber et
al. (\cite{spie}) and Granzer et al. (\cite{malaga2}).

Data reduction is performed automatically using the IRAF\footnote{The
Image Reduction and Analysis Facility is hosted by the National
Optical Astronomy Observatories in Tucson, Arizona at URL
iraf.noao.edu.}-based STELLA data-reduction pipeline (Weber et al.
\cite{spie11}). Images were corrected for bad pixels and
cosmic-ray impacts. Bias levels were removed by subtracting the
average overscan from each image followed by the subtraction of
the mean of the (already overscan subtracted) master bias frame.
The target spectra are flat fielded with a nightly master
flat which has been normalized to unity. The nightly master flat
itself is constructed from around 50 individual flats observed
during dusk, dawn, and around midnight. After removal of the
scattered light, the one-dimensional spectra were extracted with
the standard IRAF optimal extraction routine.
The blaze function was then removed from the target spectra,
followed by a wavelength calibration using consecutively recorded
Th-Ar spectra. Finally, the extracted spectral orders were
continuum normalized by dividing them with a flux-normalized
synthetic spectrum of the same spectral classification as the
target in question.

\section{Magnetic field measurement}\label{S3}

\subsection{PCA reconstruction of the Zeeman pattern}
\label{ZeemanPattern}

The principal component analysis (PCA) or Karhunen-Loeve transformation (\cite{Bishop1995}) is a multivariate algorithm, that is able to analyze tiny variations in time series. The PCA algorithm decomposes the entire data set of observed Stokes~I profiles into a new coordinate system, which is of advantage to detect the directions of high variances (see \cite{Gonzales2008} and \cite{Carroll2009}). The new basis has to explain the largest amount of variances of the data set by using as few as possible orthogonal basis vectors. The PCA computes this set of new basis vectors by calculating the eigenvectors of the covariance matrix of the data. The eigenvectors, also called eigenprofiles, are ordered by their associated eigenvalues. Therefore, the PCA will project the systematic and most coherent features to the first few eigenprofiles and will separate them from noise and from each other.

With a set of computed eigenprofiles $\vect{f}_i(\lambda)$, one can reconstruct the Stokes~I profiles $\vect{I}(t,\lambda)$ by
\begin{equation}
\vect{I}(t, \lambda)=\sum_i c_i(t)\vect{f}_i(\lambda)
\end{equation}
where $c_i(t) = \vect{I}(t, \lambda) \vect{f}_i(\lambda)$ is the scalar product between the Stokes~I profile $\vect{I}(t, \lambda)$ and the eigenprofile $\vect{f}_i(\lambda)$. The coefficient $c_i(t)$ is therefore a measure for how strong a certain feature, characterized by an eigenprofile $\vect{f}_i(\lambda)$, is present in the Stokes~I profile given at time $t$. 

\begin{figure}
\center
\includegraphics[trim=40 0 10 0, angle=0,width=80mm,clip]{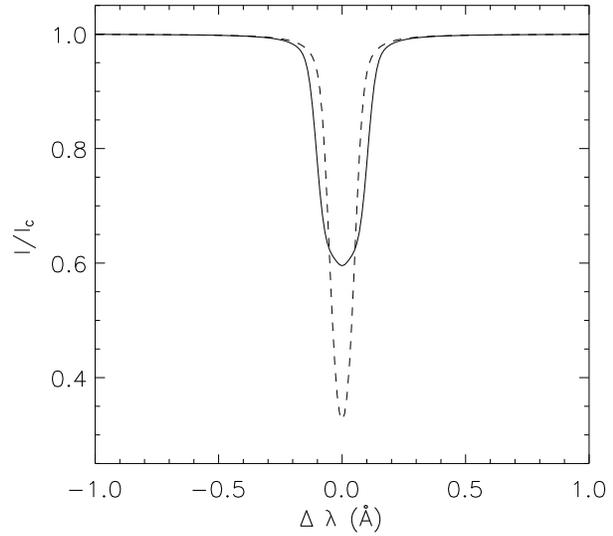}
\caption{Synthetic line profiles of the magnetic-sensitive Fe\,{\sc i} line at 6173\,\AA\ without a magnetic field (dashed line) and with a surface averaged magnetic field of 1.3\,kG (full line). }\label{F2}
\end{figure}

It is well known that the magnetic field affects spectral lines in a certain way. An example is shown in Fig.~\ref{F2}. We simulate the Fe\,{\sc i} 6173\,\AA\ line without magnetic field and with a magnetic field of 1300\,G on conditions similar to \epseri. It is clearly visible that the magnetic field leads to broadened line wings and thus decreased line depth. The variations in the activity of \epseri\ will produce tiny variations with similar shape in the spectral lines. To detect such little specific variations in the time series we used the PCA.

\cite{Skumanich2002} demonstrated the physical content of PCA applied to Stokes profiles. They used the weak field approximation, which assumes that the magnetic field is weak enough to represent the Stokes~I profiles by the low order terms of its Taylor series. 
\begin{align}
\vect{I}(t, \lambda)&=\vect{I}_0(t, \lambda) + \alpha(t)\partial\vect{I}_0(t, \lambda)/\partial \lambda\ +\nonumber \\ 
& \qquad \beta(t)\partial\vect{I}^2_0(t, \lambda)/\partial\lambda^2+ \ldots,
\end{align} 
where $\vect{I}_0(t, \lambda)=c_0(t)\vect{f}_0(\lambda)$. \cite{Skumanich2002} show that the coefficient $\alpha(t)$ could be associated with the line-of-sight velocity $v_{\rm los}$ and $\beta(t)$ with the magnetic Zeeman splitting $\Delta\lambda_{\rm Z}$. With the help of \cite{Jefferies1989} they showed that
\begin{equation}
\beta(t) \equiv (1+\cos^2 \phi) \Delta\lambda_{\rm Z}^2/2,
\label{bpropB}
\end{equation}
where $\Delta\lambda_{\rm Z} = \frac{e_0\lambda^2}{4\pi m c^2 \Delta\lambda} B(t)$, so that $\beta(t) \propto B^2(t)$.  \cite{Skumanich2002} and \cite{Jefferies1989} emphasized that the variations caused by magnetic fields will have the shape of the second derivative of the Stokes~I profiles and could be analyzed by its coefficients.

\begin{figure}
{\bf a.}\\
\includegraphics[trim=14 0 12 0, angle=0,width=80mm,clip]{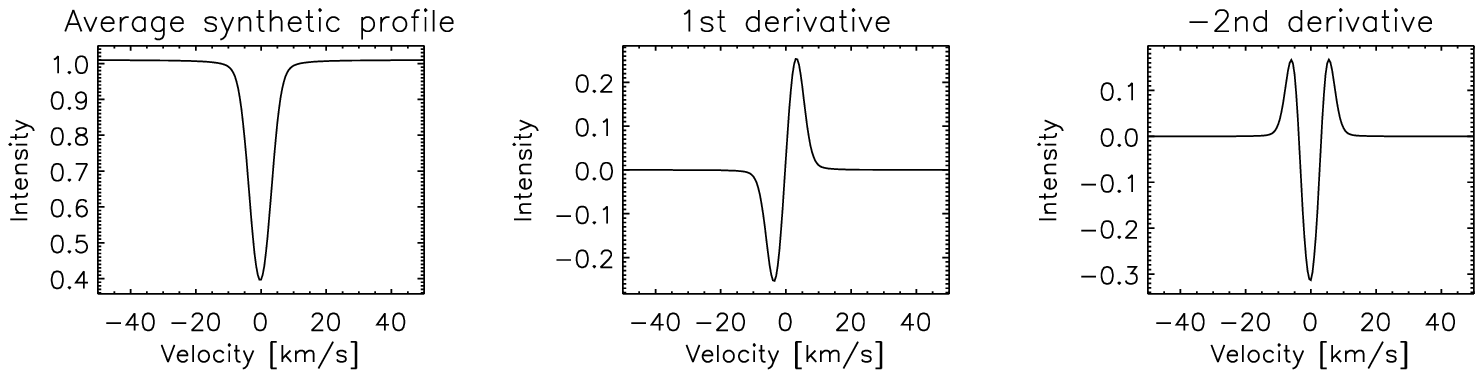}

{\bf b. }\\
\includegraphics[trim=14 0 12 0, angle=0,width=80mm,clip]{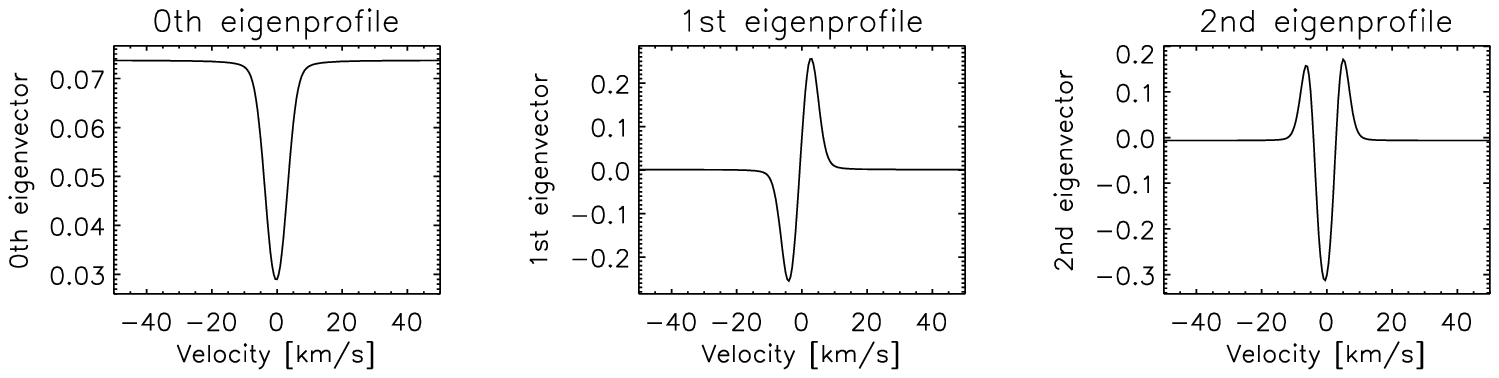}
\caption{Synthetic line profiles analysis for Fe\,{\sc i} 6173 by varying the line-of-sight velocity from 0\,--\,500\,\ms\ and the magnetic field from 0\,--\,300\,G. {\bf a} From left to right the average synthetic profile and its first and second derivative. {\bf b} The first three eigenprofiles computed with PCA.}\label{F3}
\end{figure}

To prove this prediction, we create a set of 101 synthetic Fe\,{\sc i} 6173\,\AA\ line profiles which vary in line-of-sight velocity from 0\,--\,500\,\ms\ and in the magnetic field from 0\,--\,300\,G. In Fig.~\ref{F3}a the average mean profile and its first and second derivative is shown, Fig.~\ref{F3}b (the row below) shows the first three eigenprofiles computed by the PCA. The first and second eigenprofile matches very well with the first and second derivative, as the theory predicted. The variations in velocity and magnetic field are orthogonal to each other and can be separated by PCA. Additionally, the zeroth eigenprofile shows the shape of the average synthetic line, as expected, because the greatest variation of the Stokes~I profiles to a zero-line profile is the average line profile. Therefore, the PCA algorithm enables us to detect tiny  variations caused by the magnetic field in our time series of 338 Stokes~I spectra.

%
%
%

%
%

\subsection{Selection of spectral lines and denoising}

\begin{figure}
\includegraphics[trim=40 0 10 0, angle=0,width=80mm,clip]{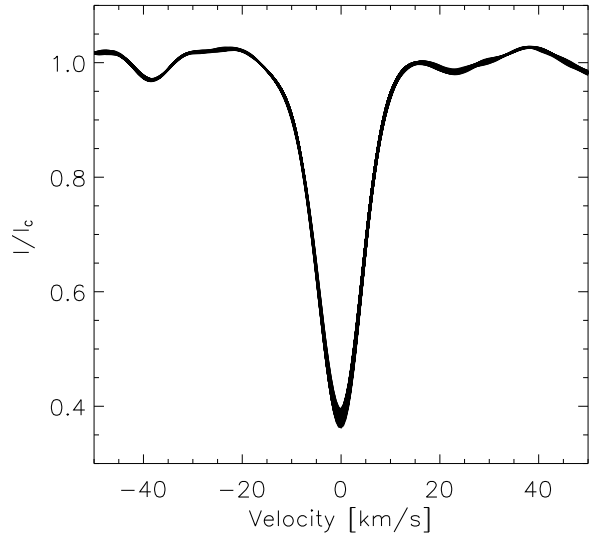}
\caption{The 338 normalized spectral lines profiles used for magnetic-field analysis with PCA.}
\label{Profiles}
\end{figure}

Spectral lines are selected from the VALD-II database (Kupka et al.~\cite{vald2}). The selection focused on lines of approximately equal strength, as unblended as possible, and with a high Land\'e factor. As threshold we set 0.8 in the line depth and a minimum Land\'e factor of 1.59. A total of 30 such lines were selected and are listed in Table~\ref{T1}. For these lines we expect a comparable dependency on other line-broadening mechanisms. All 30 individual spectral lines were transformed into velocity space and averaged to obtain mean spectral lines profiles. Therefore, the S/N-ratio increased by approximately a factor of five. In Fig.~\ref{Profiles}, we show the 338 normalized spectral lines profiles we used as our Stokes~I profiles for analysis.
\begin{table}
\caption{Spectral lines used in this analysis. } \label{T1}
\begin{tabular}{lll||lll}
\hline \noalign{\smallskip}
El. & $\lambda$ (\AA ) & Land\'e  & El. & $\lambda$ (\AA ) & Land\'e  \\
         &  & factor &          &  & factor\\
\hline
Cr\,{\sc i} & 4545.953 & 1.92 & Fe\,{\sc i} & 4903.310 & 2.24\\
Cr\,{\sc i} & 4591.391 & 2.00 & Fe\,{\sc ii} & 4923.927 & 1.69\\
Fe\,{\sc i} & 4602.941 & 1.74 & Fe\,{\sc i} & 4938.814 & 2.01\\
Fe\,{\sc i} & 4611.279 & 1.89 & Cr\,{\sc i} & 4942.496 & 1.85\\
Cr\,{\sc i} & 4616.124 & 1.67 & Cr\,{\sc i} & 4964.927 & 2.17\\
Cr\,{\sc i} & 4626.173 & 2.01 & Fe\,{\sc i} & 4967.897 & 1.88\\
Fe\,{\sc i} & 4635.846 & 2.09 & Fe\,{\sc i} & 5044.211 & 1.77\\
Fe\,{\sc i} & 4673.270 & 1.84 & Cr\,{\sc i} & 5348.315 & 1.59\\
Fe\,{\sc i} & 4690.138 & 2.16 & Mn\,{\sc i} & 5394.677 & 1.86\\
Fe\,{\sc i} & 4704.948 & 2.49 & Mn\,{\sc i} & 5432.546 & 2.14\\
Mn\,{\sc i} & 4754.042 & 1.65 & Fe\,{\sc i} & 5497.516 & 2.26\\
Fe\,{\sc i} & 4768.396 & 1.62 & Fe\,{\sc i} & 5501.465 & 1.88\\
Mn\,{\sc i} & 4783.427 & 1.97 & Fe\,{\sc i} & 5506.779 & 2.00\\
Fe\,{\sc i} & 4848.883 & 2.01 & Fe\,{\sc i} & 6213.430 & 2.00\\
Fe\,{\sc i} & 4878.211 & 3.00 & Fe\,{\sc i} & 6219.281 & 1.66\\
\noalign{\smallskip}
\hline
\end{tabular}
\end{table}

%
%
%

\subsection{Eigenprofiles analysis}

To make the PCA more sensitive to small variations in the wings of the spectral lines profiles, we analyzed the inverted ($1-$\,Stokes~I) profiles.

\begin{figure*}
{\bf a.}\\
\includegraphics[trim=12 0 8 0, angle=0,width=160mm,clip]{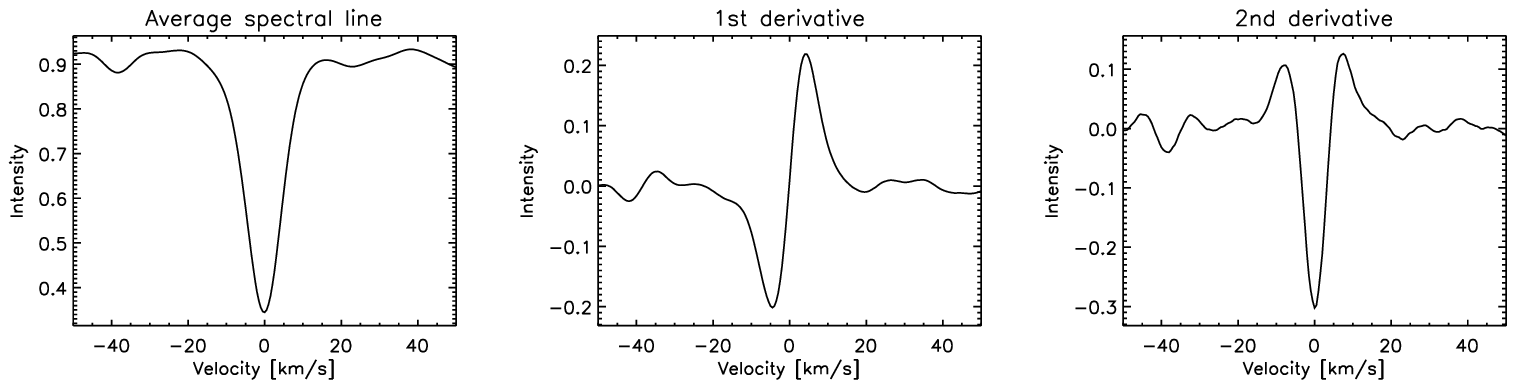}

{\bf b.}\\
\includegraphics[trim=12 0 8 0, angle=0,width=160mm,clip]{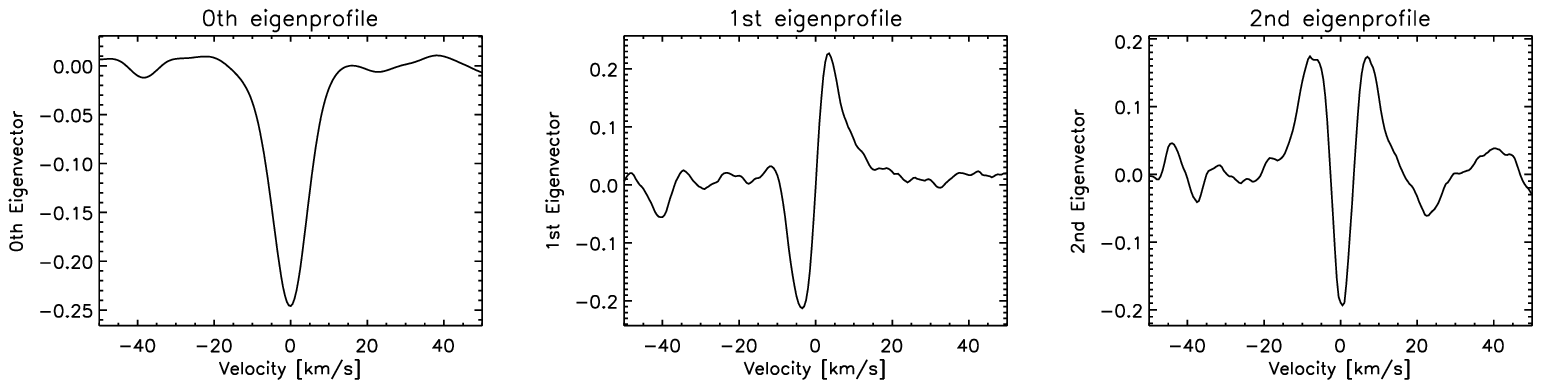}
\caption{Comparison of theoretically expected profile patterns with results from PCA. {\bf a} The observed averaged spectral lines profile and its first and second derivative. {\bf b} The first three eigenprofiles of the spectral lines profiles.}
\label{F5}
\end{figure*}

Fig.~\ref{F5} shows the comparison of the theoretical expectations (see Sec.~\ref{ZeemanPattern}) with  our analysis results. The first row, Fig.~\ref{F5}a, shows the average profile over all spectra in the left panel, and its first and second derivative in the other panels. It illustrates the pattern we are looking for. The second row, Fig.~\ref{F5}b, shows the zeroth, first, and second eigenprofile of the spectral lines profiles derived with PCA. The zeroth eigenprofile in the left column represents the mean profile of the 338 spectral lines profiles. As expected, the variations of the second eigenprofile match the second derivative of the observed averaged profile. The variations are in good agreement with a correlation coefficient of 82\,\%. This correlation indicates, following \cite{Skumanich2002}, variations in the magnetic field affecting the spectral lines on \epseri. By comparison, the first eigenprofile bears resemblance with the first derivative. According to Sect.~\ref{ZeemanPattern} this eigenprofile indicates a velocity drift of the 338 spectral lines profiles against each other and were not further examined. The first and second eigenprofiles  are orthogonal to each other, so that the effects from the velocity drift are completely separated from signals resulting from magnetic field variations. This proves that the variations detected reside mostly in the second eigenprofile due to the joint Zeeman pattern. The eigenprofiles in higher orders are containing noise. 

We note that in our PCA all available spectral line profiles of different wavelengths and times have to
enter into the analysis simultaneously in order to achieve the disentangling of the individual eigenprofiles into
patterns which represent the first and the second spectral derivative (and thus a velocity and magnetic field
variation). We experimented with two subsets of the given set of spectral lines, i.e., we separated the very high
Land\'e factor lines ($\geq$2.0) from the low and moderate Land\'e factors ($<$2.0) and performed the same analysis on
the two data sets as before. Neither of the two data sets could reconstruct an eigenprofile which matches the
second derivative with an acceptable correlation, nor could such an eigenprofile be found within the most
significant ones. This is not a problem of the new technique but of the data quality, the split data simply do not
allow such a low detection threshold anymore.

%
%
%

%
%
%
\begin{figure}
\includegraphics[angle=0,width=80mm,clip]{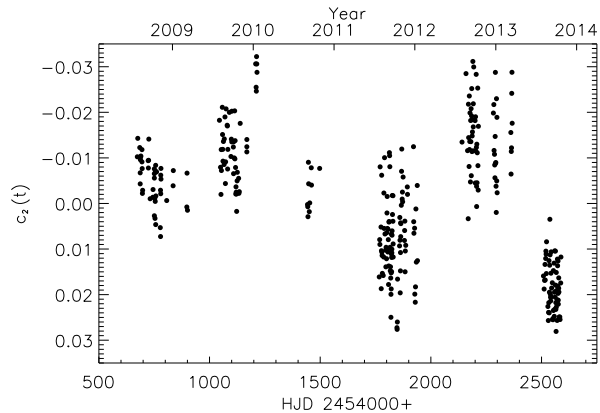}
\caption{The coefficients of the second eigenprofile of the 338 spectral lines profiles derived with PCA.}
\label{Coeff}
\end{figure}

The coefficients of the second eigenprofile $c_2 (t)$, shown in Fig.~\ref{Coeff}, can now be used to quantify the magnetic field and its variations. A coefficient of $c_2 (t) \approx 0$ represents the average magnetic field at time $t$, while positive and negative coefficients represent accordingly weaker or stronger fields. The coefficients in Fig.~\ref{Coeff} show roughly a long-term oscillation around zero. We also see scattering of the coefficients on small time scales. The single seasons are clearly separated by different seasonal mean coefficients. In the next step, we will convert the coefficients to values of magnetic field.

\subsection{Magnetic field determination}

%
%
%
\begin{figure}
\includegraphics[angle=0,width=80mm,clip]{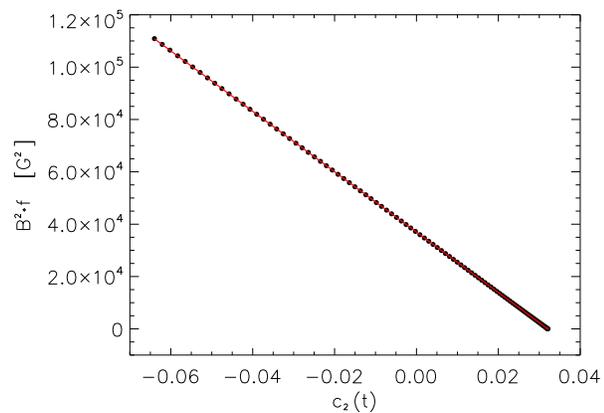}
\caption{The calibration curve used to convert coefficients of the second eigenprofile $c_2 (t)$ to magnetic field values $B^2\cdot f$. Every tenth coefficient of the 1,000 synthetic profiles is plotted and the corresponding linear fit function.}
\label{Calb}
\end{figure}

The actual quantification of the magnetic field, is done through a least-squares comparison of the coefficients from the observations with synthetic profiles of known \Bf. We assumed $f=1$ to obtain a surface average magnetic field and generated 1,000 synthetic profiles with \Bf\, between 0\,G and 333\,G to analyze this series with PCA. The resulting $c_2 (t)$ coefficients relate linearly with $(B\cdot \sqrt{f})^2$, see Eq. (\ref{bpropB}), and a simple regression then defines the transformation to surface-averaged magnetic field strength. 
With the help of the calibration curve, Fig.~\ref{Calb}, we convert the coefficients of the second eigenprofile $c_2 (t)$ to magnetic field values \Bf. Hereinafter we defined for the surface average magnetic field
\begin{equation}
B \cdot \sqrt{f}_{f=1} = \langle B \rangle. 
\end{equation}

%
%
%
\begin{figure*}
\includegraphics[trim=12 0 10 0, angle=0,width=\textwidth,clip]{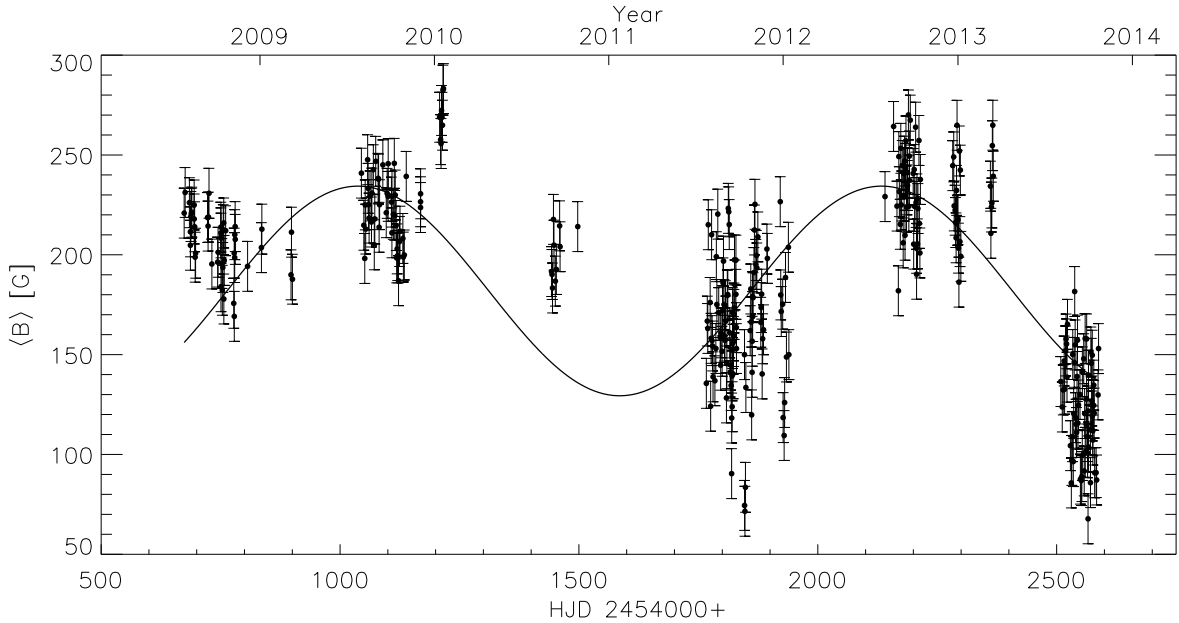}
\caption{Surface averaged magnetic field measures for the observing seasons 2008/09 to 2013/14. The solid line
indicates a sine-fit using a period of 1096\,d ($\approx$3.0\,yr), derived by a Lomb-Scargle periodogram.}
\label{MagField}
\end{figure*}

\section{Results}\label{S4}

\subsection{The magnetic field of \epseri}

Fig.~\ref{MagField} shows all 338 individual \B\ measures with its uncertainties. We measure a mean magnetic field
over all seasons of 186$\pm$47\,G, where the error represents the standard deviation. The mean magnetic fields of
single seasons are listed in Table~\ref{T3}. We see that the average magnetic field of the single seasons is
oscillating with maxima in the seasons 2009/2010 and 2012/2013 and minima in 2011/2012 and 2013/2014. The
variations in single seasons are largest in the seasons 2011/2012 and 2012/2013 and smallest in the season
2013/2014. In the very end of season 2009/2010, we clearly see a separated group of few observations where magnetic
field values are higher as in the rest of the season. Season 2010/2011 consists of only 10 observations, which
makes the variation analysis not reliable, but the data points fit in the overall trend.

Another issue from Fig.~\ref{MagField} is that the magnetic field values vary over few tens of Gauss at significantly shorter time scales then the rotation period of \epseri . Their average internal errors are $\pm$13\,G. Short-term variations on a 1\,--\,5\,$\sigma$ level are evident within an observing season. This could be an indication that the topology of \epseri 's magnetic field is spatially inhomogeneous. The strong variations could appear from local active regions which rotate in and out of the line of sight. This would be in a good agreement to the large-scale magnetic field geometry determined by \cite{Jeffers2014}. A periodogram analysis does not reveal a significant short period, which could be interpreted as an indicator for temporal inhomogeneities of the active regions.


%
%
\begin{table}
\caption{Seasonal values of magnetic field measurements.} \label{T3}
\begin{tabular}{ccc}
\hline \noalign{\smallskip}
Season & Number of & Mean magnetic field \\
         & observations & \B\ (G)  \\
\noalign{\smallskip}\hline \noalign{\smallskip}
2008/2009 & 45 & $206 \pm 15$  \\
2009/2010 & 55 & $229 \pm 23$  \\
2010/2011 & 10 & $200 \pm 13$  \\
2011/2012 & 98 & $165 \pm 32$  \\
2012/2013 & 65 & $230 \pm 21$  \\
2013/2014 & 65 & $124 \pm 25$  \\
\noalign{\smallskip}\hline \noalign{\smallskip}

all & 338 & $186 \pm 47$  \\     
\noalign{\smallskip}\hline
\end{tabular}
\end{table}

Temperature surface inhomogeneities like starspots may contribute to the disk-integrated line strength, and thus
its broadening, even if the spots remain unresolved. Because in the present paper it is our intention (like in most
other Zeeman-broadening works) to determine the surface averaged magnetic field, one of the built-in limitations is
the still common (but surely simplified) homogeneous two-component photospheric model, i.e. two magnetic components
described by a single temperature model (see also Sect.~6.4 in Valenti et al. (\cite{val}) for a detailed
discussion of the possible error budget of this assumption).

\subsection{Time variability}

\begin{figure}
\includegraphics[angle=0,width=80mm,clip]{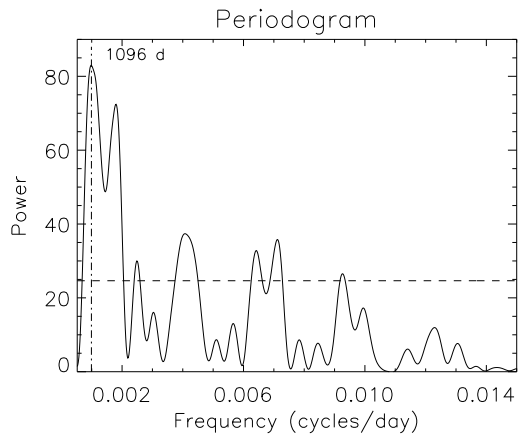}
\caption{Lomb-Scargle periodogram for Fig.~\ref{MagField}. The dashed line represents a FAP of 10$^{-5}$. The
vertical dashed-dotted line indicates the maximum peak.}
\label{MagFieldPeriodogram}
\end{figure}

Fig.~\ref{MagField} shows a sine-curve fit to the magnetic-field data that indicate a period of 3.0\,yr (solid line). In Fig.~\ref{MagFieldPeriodogram} the corresponding periodogram is given.

\cite{Hatzes2000} were the first to report a 3-years magnetic activity cycle in \epseri. Recently \cite{Metcalfe2013} suggested the simultaneous existence of two cycle periods of 2.95$\pm$0.03\,years and 12.7$\pm$0.3\,years. They used the Ca\,{\sc ii} H\&K S-index to estimate the magnetic activity cycles. Our time series of \B\ measurements of \epseri\ show a similar behavior as the published data of the Ca\,{\sc ii} H\&K S-index in Figure~1 by \cite{Metcalfe2013} and the published data of the S-index in Figure~2 by \cite{Jeffers2014}.

We analyzed our time series for long-term periods with the help of a Lomb-Scargle \cite{Scargle1982} analysis and determine a period of 1096$\pm$471\,d (3.0$\pm$1.3\,yr). To estimate its false alarm probability (FAP) we used two different methods. In the first one we apply the formula of \cite{Horne1986} which resulted in a FAP well
below $10^{-7}$ for the most significant peak. For a second method, we used a bootstrap resampling technique where we randomly redistributed the eigenvalue coefficients but retained the same times for the observations (e.g., K\"urster et al. \cite{Kuerster97}). A 100,000 resampling trials were performed but none had reached a power as high as our most significant peak and we estimate the FAP to be lower than 10$^{-5}$. 

%
%
%

The sine-curve periodicity in Fig. \ref{MagField} is comparable to the trend of the Ca\,{\sc ii} H\&K S-index in Figure~1 of \cite{Metcalfe2013}. We see in our data and in the S-index measurements that the activity of \epseri\ is higher in 2009/2010 and lower in 2008/2009 and
2011/2012. The rise of activity in the year 2012/2013 is suggestively visible by \cite{Metcalfe2013}. Also the
higher variances in the years 2011/2012 and 2012/2013 are visible in the Ca\,{\sc ii} H\&K measurements as well.
Good agreement is also seen for  the high activity at the end of the season 2009/2010. The overall trend of our
results also fits with the S-index measurements from \cite{Jeffers2014} shown in their Figure~2. The behavior of the data
from 2009/2010 to 2012/2013 is similar to our results of the \B\ measurements. The last season (2013/2014) seems to
be less active than the seasons before. The trend of decreasing activity in 2013/2014 is also in agreement with the
S-index measurements of \cite{Jeffers2014}.

\section{Conclusions}\label{S5}

Using PCA on a time series of 338 spectra, we were able to quantify the magnetic activity variations
on \epseri. We computed an average surface-mean magnetic field of 186$\pm$47\,G. The analysis showed
variations of the mean surface magnetic field on shorter time scales as the rotation period, which
could be interpreted as spatial inhomogeneities in the magnetic field. The lack of a long-term
persistent pattern on time scales of the rotation period suggests additional temporal
inhomogeneities in the magnetic activity of \epseri.

We confirm the qualitative agreement with the Ca\,{\sc ii} H\&K S-index analysis done by \cite{Metcalfe2013} and \cite{Jeffers2014}. The trend of the Ca\,{\sc ii} H\&K measurements are comparable to our \B\ measurements. Our analysis shows a time variation of the magnetic field that is comparable to the shorter magnetic activity cycle estimate by \cite{Metcalfe2013}. We suggest a period of 3.0$\pm$1.3\,yr comparable to the 2.95$\pm$0.03\,years determined by \cite{Metcalfe2013}. By using longer time series spectra of \epseri\ the method would principally be able to capture the longer period as well, if existent.

The analysis of time series by using PCA is a very successful tool to detect variations on cool, slowly rotating stars. The velocity variations can be separated from the magnetic field variations for a detailed analysis of time variations of the surface mean magnetic field. The method promises to be successful on further stars and will be used for analyzing variations of the average surface magnetic field on suitable targets in combination with the recently presented orthogonal matching-pursuit technique for single spectra by Carroll \& Strassmeier (\cite{car:str}). 

\acknowledgements The STELLA project was funded by the Science and Culture Ministry of the German State of Brandenburg (MWFK) and the German Federal Ministry for Education
and Research (BMBF). Its a pleasure to thank Michi Weber and Thomas Granzer for their efforts to keep STELLA running and an anonymous referee for many helpful suggestions.



\begin{thebibliography}{}

\bibitem[2012]{ang:but}
Anglada-Escud\'e, G., \& Butler, R. P. 2012, ApJS, 200, 15

\bibitem[2012]{bai:arm}
Baines, E. K., \& Armstrong, J. T. 2012, ApJ, 748, 72

\bibitem[1988]{bas:mar}
Basri, G., \& Marcy, G. W. 1988, ApJ, 330, 274

\bibitem[2006]{ben:mca}
Benedict, G. F., McArthur, B. E., Gatewood, G., et al. 2006, AJ, 132, 2206

\bibitem[Bishop 1995]{Bishop1995}
Bishop, C.~M.\ 1995, Neutral Networks for Pattern Recognition, Oxford University Press

\bibitem[1988]{camp}
Campbell, B., Walker, G. A. H., \& Yang, S. 1988, ApJ, 331, 902

\bibitem[Carroll et al. (2009)]{Carroll2009}
Carroll, T. A., Kopf, M., Strassmeier, K. G., \& Ilyin, I. 2009, in IAU Symp. 259, \emph{Cosmic Magnetic Fields: From Planets, to Stars and Galaxies}, Cambridge Univ. Press, p.633

\bibitem[2014]{car:str}
Carroll, T. A., \& Strassmeier, K. G. 2014, A\&A, 563, A56
 
\bibitem[2006]{cro:wal}
Croll, B., Walker, G. A. H., Kuschnig, R., et al. 2006, ApJ, 648, 607

\bibitem[1996]{don:saa}
Donahue, R. A., Saar, S. H., \& Baliunas, S. L. 1996, ApJ, 466, 384

\bibitem[1993]{dra:smi}
Drake, J. J., \& Smith, G. 1993, ApJ, 412, 797

\bibitem[1991]{frey}
Frey, G. J., Hall, D. S., Mattingly, P., Robb, S., Wood, J., Zeigler, K., \& Grim, B.
1991, AJ, 102, 1813

\bibitem[2007]{fro}
Fr\"ohlich, H. 2007, AN, 328, 1037

\bibitem[2010]{malaga2}
Granzer, T., Weber, M., \& Strassmeier, K. G. 2010, Adv. in Astr. 2010, ID~980182

\bibitem[1995]{gray84}
Gray, D. F. 1984, ApJ 277, 640

\bibitem[1995]{gra:bal}
Gray, D. F., \& Baliunas, S. L. 1995, ApJ, 441, 436

\bibitem[Hatzes et al. (2000)]{Hatzes2000}
Hatzes, A.~P., Cochran, W.~D., McArthur, B., et al.\ 2000, ApJ, 544, L145

\bibitem[Horne \& Baliunas (1986)]{Horne1986} 
Horne, J.~H., \& Baliunas, S.~L.\ 1986, \apj, 302, 757

\bibitem[Jefferies et al. (1989)]{Jefferies1989}
Jefferies, J., Lites, B.~W., \& Skumanich, A.\ 1989, \apj, 343, 920 

\bibitem[Jeffers et al. (2014)]{Jeffers2014}
Jeffers, S.~V., Petit, P., Marsden, S.~C., et al.\ 2014, \aap, 569, A79 

\bibitem[2011]{koch}
Kochukhov, O., Makaganiuk, V., Piskunov, N. et al. 2011, ApJ 732, L19

\bibitem[1999]{vald2}
Kupka, F., Piskunov, N., Ryabchikova, T. A., Stempels, H. C., \& Weiss, W. W. 1999,
A\&AS, 138, 119

\bibitem[1997]{Kuerster97} 
K\"urster, M., Schmitt, J.~H.~M.~M., Cutispoto, G., \& Dennerl, K.\ 1997, \aap, 320, 831

\bibitem[1984]{marcy}
Marcy, G. W. 1984, ApJ, 276, 286

\bibitem[1989]{mar:bas}
Marcy, G. W., \& Basri, G. 1989, ApJ, 345, 480

\bibitem[Mart{\'{\i}}nez Gonz{\'a}lez et al. (2008)]{Gonzales2008} 
Mart{\'{\i}}nez Gonz{\'a}lez, M.~J., Asensio Ramos, A., Carroll, T.~A., et al.\ 2008, A\&A, 486, 637 

\bibitem[Metcalfe et al. (2013)]{Metcalfe2013} 
Metcalfe, T.~S., Buccino, A.~P., Brown, B.~P., et al.\ 2013, ApJ, 763, L26 

\bibitem[2011]{ref:qui}
Reffert, S., \& Quirrenbach, A. 2011, A\&A, 527, A140

\bibitem[1997]{ruedi}
R\"uedi, I., Solanki, S. K., Mathys, G., \& Saar, S. H. 1997, A\&A, 318, 429

\bibitem[1988]{saar}
Saar, S. H. 1988, ApJ, 324, 441

\bibitem[2009]{sav}
Savanov, I. S. 2009, Astronomy Reports, 53, 950

\bibitem[(Scargle 1982)]{Scargle1982}
Scargle, J.~D.\ 1982, \apj, 263, 835 

\bibitem[Skumanich \& L{\'o}pez Ariste (2002)]{Skumanich2002}
Skumanich, A., \& L{\'o}pez Ariste, A.\ 2002, \apj, 570, 379 

\bibitem[2002]{sta:mur}
Starck, J.-L., \& Murtagh, F. 2002, Handbook of Astronomical Data Analysis, Springer, Berlin

\bibitem[1981]{ste:hol}
Steenbock, W., \& Holweger, H. 1981, A\&A, 99, 192

\bibitem[2004]{stella}
Strassmeier, K. G., Granzer, T., Weber, M., et al. 2004, AN, 325,
527

\bibitem[2010]{malaga}
Strassmeier, K. G., Granzer, T., \& Weber, M. 2010, Adv. in Astr. 2010, ID~970306

\bibitem[1995]{val}
Valenti, J. A., Marcy, G. W., \& Basri, G. 1995, ApJ, 439, 939

\bibitem[2011]{spie11}
Weber, M., Granzer, T., Strassmeier, K. G., \& Woche, M. 2011, Proc. SPIE 7019, 70190L

\bibitem[2012]{spie}
Weber, M., Granzer, T., Strassmeier, K. G., \& Woche, M. 2012, Proc. SPIE 8451, 8451-19

\bibitem[2013]{zech}
Zechmeister, M., K\"urster, M., Endl, M., et al. 2013, A\&A, 552, A78


\end{thebibliography}
\end{document}